\documentclass[fleqn,usenatbib]{mnras}

\usepackage{graphicx}
\usepackage{amsmath}
\usepackage{amssymb}
\usepackage{nicefrac}
\usepackage{multirow}
\usepackage{threeparttable}
\usepackage{footmisc}

\newcommand{\hunit}[1]{\, h^{-1} #1}
\newcommand{\solarmass}{\, \text{M}_\odot}

\defcitealias{2016MNRAS.462..250M}{M16}

\title[Producing hybrid dark matter simulations]{An efficient hybrid method to produce high resolution large volume dark matter simulations for semi-analytic models of reionisation}

\author[Qiu et al.]{Yisheng Qiu$^{1,3}$\thanks{E-mail: yishengq@student.unimelb.edu.au}, Simon J. Mutch$^{1,3}$, Pascal J. Elahi$^{2,3}$, Rhys J. J. Poulton$^{2,3}$, \newauthor Chris Power$^{2,3}$, J. Stuart B. Wyithe$^{1,3}$
\\$^{1}$School of Physics, University of Melbourne, Parkville, VIC 3010, Australia
\\$^{2}$International Centre for Radio Astronomy Research, M468, University of Western Australia, 35 Stirling Hwy, Perth, WA 6009, Australia
\\$^{3}$ARC Centre of Excellence for All Sky Astrophysics in 3 Dimensions (ASTRO 3D)
}

\date{Accepted XXX. Received YYY; in original form ZZZ}
\pubyear{2020}

\begin{document}
\label{firstpage}
\pagerange{\pageref{firstpage}--\pageref{lastpage}}
\maketitle

\begin{abstract}
Resolving faint galaxies in large volumes is critical for accurate cosmic reionisation simulations. While less demanding than hydrodynamical simulations, semi-analytic reionisation models still require very large N-body simulations in order to resolve the atomic cooling limit across the whole reionisation history within box sizes $\gtrsim 100 \hunit{\rm Mpc}$. To facilitate this, we extend the mass resolution of N-body simulations using a Monte Carlo algorithm. We also propose a method to evolve positions of Monte Carlo halos, which can be an input for semi-analytic reionisation models. To illustrate, we present an extended halo catalogue that reaches a mass resolution of $M_\text{halo} = 3.2 \times 10^7 \hunit{\solarmass}$ in a $105 \hunit{\rm Mpc}$ box, equivalent to an N-body simulation with $\sim 6800^3$ particles. The resulting halo mass function agrees with smaller volume N-body simulations with higher resolution. Our results also produce consistent two-point correlation functions with analytic halo bias predictions. The extended halo catalogues are applied to the \textsc{meraxes} semi-analytic reionisation model, which improves the predictions on stellar mass functions, star formation rate densities and volume-weighted neutral fractions. Comparison of high resolution large volume simulations with both small volume or low resolution simulations confirms that both low resolution and small volume simulations lead to reionisation ending too rapidly. Lingering discrepancies between the star formation rate functions predicted with and without our extensions can be traced to the uncertain contribution of satellite galaxies.
\end{abstract}


\section{Introduction}
Simulating the epoch of reionisation is extremely challenging, with different techniques developed to study different aspects of the problem. For example, high resolution hydrodynamical simulations \citep[e.g.][]{2012ApJ...745...50W,2013MNRAS.428.1857J,2017MNRAS.470.2791C,2018MNRAS.479..994R} can resolve the faintest galaxies with detailed spatial information on the interstellar media (ISM). These faint sources are found to have non-negligible contributions to reionisation \citep{2014MNRAS.442.2560W,2020MNRAS.tmp..591K}. However, these simulations are limited to a small volume ($\lesssim 10^3 \, h^{-3} \text{Mpc}^3$). At the other extreme, \cite{2014MNRAS.439..725I} presented a study in a $425 \hunit{\rm Mpc}$ box, and pointed out that at least a $\sim 100 \hunit{\rm Mpc}$ box is required for the convergence of reionisation histories. Other studies use semi-numerical calculations of reionisation to simulate large volumes \citep[e.g][]{2015MNRAS.449.4246G,2016MNRAS.457.1550H,2019MNRAS.484..933P}. A disadvantage of these approaches is the absence of a detailed galaxy formation model. Whilst large volumes have been achieved by several hydrodynamical simulations \citep[e.g.][]{2016MNRAS.455.2778F,2018MNRAS.475..648P}, they cannot resolve the faintest sources. The Cosmic Reionisation on Computers project \citep{2014ApJ...793...29G,2014ApJ...793...30G,2015ApJ...810..154K} aims to produce hydrodynamical simulations with both large volume and high spatial resolution, with self-consistent treatment of radiative transfer, gas dynamics and star formation. They reach a $\sim 100$ pc spatial resolution in a $\sim 80 \hunit{\rm Mpc}$ box. However, one shortcoming of hydrodynamical simulations is that they are extremely computationally expensive, and therefore cannot be easily used to explore different model variations. 
\par
Semi-analytic galaxy formation models \citep[see][for reviews]{2006RPPh...69.3101B,2015ARA&A..53...51S} provide a good alternative, and can potentially achieve very high mass resolution in large volumes. They take halo merger trees extracted from N-body simulations as an input, and evolve several key baryonic components of galaxies within these halos. They do not consider hydrodynamic forces or the spatial distribution of the ISM, which limits their predictive power but makes them computationally efficient. One example is the \textsc{meraxes} semi-analytic model \citep{2016MNRAS.462..250M}, which couples galaxy formation with reionisation using \textsc{21cmfast}  \citep{2007ApJ...669..663M}. Predictions for reionisation using \textsc{meraxes} can be found in \cite{2016MNRAS.462..804G}.
\par
The mass resolution and the simulation volume of semi-analytic models are determined by the input N-body simulations. Predictions of cosmic reionisation may require a volume greater than $100^3 \, h^{-3} \text{Mpc}^3$. For example, \cite{2020arXiv200406709D} suggested that a $170 \hunit{\rm Mpc}$ box is needed for a simulation to predict convergent 21cm power spectra. At the same time, the main contribution of ionising photons could be from faint sources  (\citealp[e.g.][]{2016MNRAS.462..235L,2019ApJ...879...36F,2020MNRAS.tmp..591K}, however see \citealp{2020ApJ...892..109N}). In order to resolve all faint sources and examine their contribution to reionisation, semi-analytic models require N-body simulations with a very large particle number. This work attempts to overcome this challenging task by augmenting N-body halo merger trees using Monte Carlo halos. The first such method was presented in \cite{2016ComAC...3....3B}. We extend their study to $z \geq 5$, and introduce an improvement to make the results satisfy the halo mass function of the given N-body simulation. Detailed reionisation calculations require the spatial distribution of halos. This work also proposes an approach to assign and evolve the position of Monte Carlo halos, which can reproduce halo clustering predicted by the N-body simulation.
\par
This paper is organised as follows. Our methodology of extending N-body halo catalogues is presented in Section \ref{sec:method}. Specifically, Section \ref{sec:sim} describes the N-body simulations utilised in this work. Section \ref{sec:algorithm} introduces the algorithms to augment N-body halo merger trees. We populate and evolve the position of Monte Carlo halos in Section \ref{sec:pos}, and sample their spin parameter in Section \ref{sec:spin}. Then, in Section \ref{sec:app}, we apply the extended halo catalogues to the \textsc{meraxes} semi-analytic reionisation model. Finally, this work is summarised in Section \ref{sec:summary}.

\section{Methodology} \label{sec:method}
\subsection{N-body simulations} \label{sec:sim}
This work utilises two boxes from the Genesis N-body simulations (Elahi et al., in preparation). We focus on extending the mass resolution of L105N2048, which is a $105 \hunit{\rm Mpc}$ box, containing $2048^3$ particles, with $m_\text{p} = 1.17 \times 10^7 \, \hunit{\solarmass}$. To calibrate and verify our results, we take advantage of L35N2650, which has a much higher resolution. It contains $2650^3$ particles in a $35 \hunit{\rm Mpc}$ box. The particle mass is $m_\text{p} = 2.00 \times 10^5 \hunit{\solarmass}$. All the simulations are run using \textsc{gadget-2} \citep{2005MNRAS.364.1105S}. Halos in the simulations are identified using \textsc{velociraptor} \citep{2019PASA...36...21E,2019ascl.soft11020E}, which is a  six-dimensional friends-of-friends phase space halo finder. Merger trees are constructed using \textsc{treefrog} \citep{2019PASA...36...28E,2019ascl.soft11021E}. Table \ref{table:cat} provides a summary of the N-body halo catalogues used in this work. Throughout the paper, we adopt the mass obtained by summing all particles in a friends-of-friends group as halo mass. The Genesis N-body simulations use a cosmology with $h = 0.6751$, $\Omega_\text{m} = 0.3121$, $\Omega_\text{b} = 0.0491$, $\Omega_\Lambda = 0.6879$, $\sigma_8=0.8150$, $n_\text{s} = 0.9653$ \citep[fourth column in Table 4 of][]{2016A&A...594A..13P}. To be consistent, we adopt this cosmology throughout paper.

\subsection{Augmenting N-body merger trees} \label{sec:algorithm}
Our approach to augment N-body merger trees mainly follows \cite{2016ComAC...3....3B}. The basic idea is to generate Monte Carlo merger trees with the desired mass resolution and compare these with an N-body merger tree in the mass range where the simulation is fully reliable. If both trees are similar, as determined by several criteria (described below), Monte Carlo halos with mass below the simulation resolution are attached to the N-body merger tree. This results in a hybrid structure, containing both Monte Carlo and N-body halos, but with the same mass resolution as the Monte Carlo tree.

\begin{table} 
    \centering
	\caption{Parameters of the Monte Carlo tree algorithm.}
    \label{table:params_mc}
    \begin{tabular}{c|c|c}
	\hline \hline
    Symbol & \cite{2008MNRAS.383..557P} & This work \\
	\hline
    $G_0$      & 0.57 & 1.0 \\
    $\gamma_1$ & 0.38 & 0.2 \\
    $\gamma_2$ & -0.01 & -0.4 \\
    \hline
    \end{tabular}
\end{table}

\begin{figure*}
	\includegraphics[width=\textwidth]{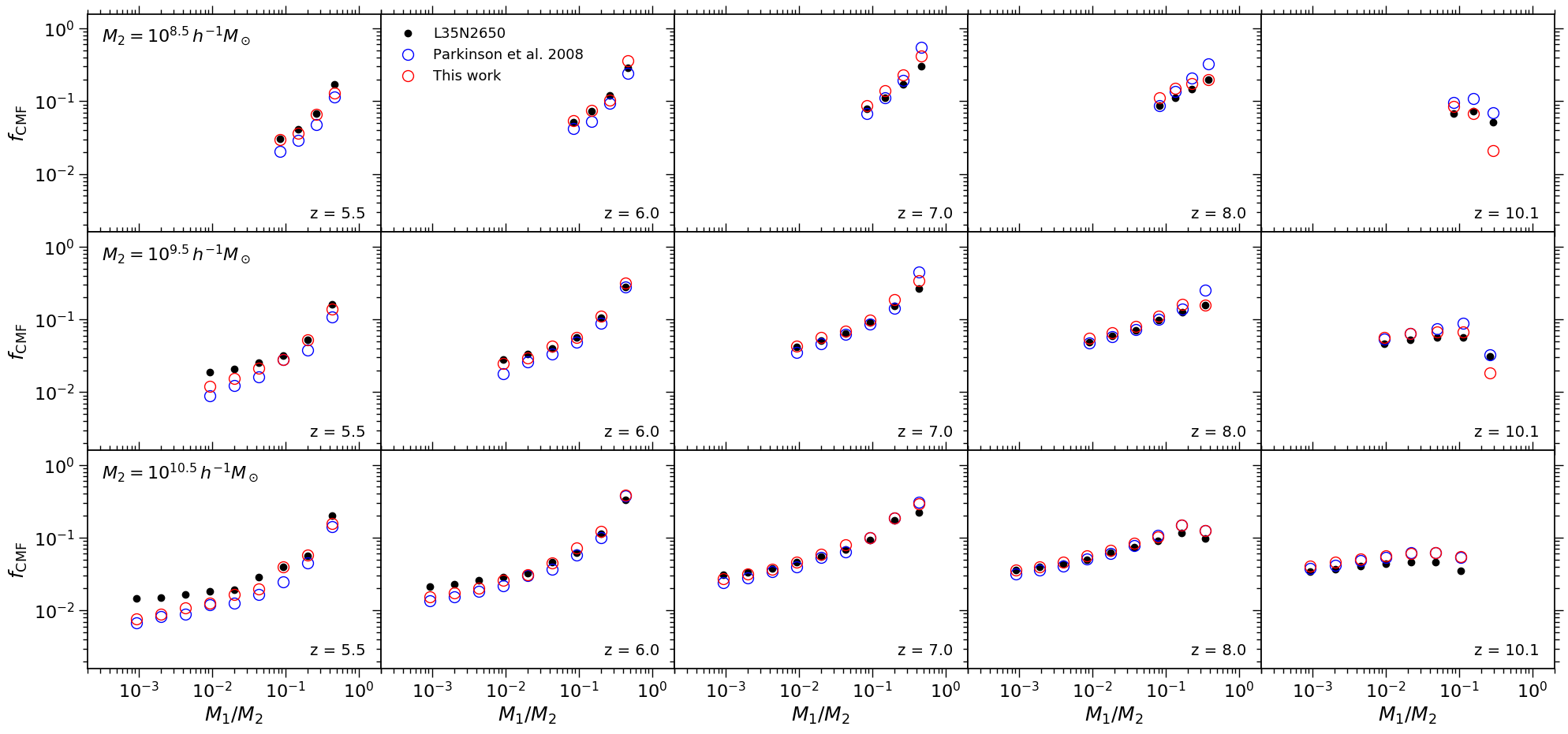}
    \caption{Fitting results of the calibration for the \citet{2008MNRAS.383..557P} algorithm. The conditional mass functions are defined by $d f_\text{CMF} / d \ln M_1$. Black dots are the fitting data, which are estimated using L35N2650. Red and blue empty circles are the results corresponding to the best-fit parameters obtained in this work and those used by \citet{2008MNRAS.383..557P} respectively. The values of the parameters are listed in Table \ref{table:params_mc}.}
    \label{fig:fit}
\end{figure*}

\subsubsection{Generating Monte Carlo trees} \label{sec:mc_tree}
We adopt the \cite{2008MNRAS.383..557P} algorithm to generate Monte Carlo merger trees. The algorithm is based on binary splits in small internal time steps. It employs the conditional mass function \footnote{The conditional mass function discussed here is defined by the mass fraction distribution ($M_1/M_2$) as a function of progenitor mass $M_1$ given the descendant mass $M_2$.} derived from the Extended Press Schechter (EPS) theory \citep{1991MNRAS.248..332B,1991ApJ...379..440B,1993MNRAS.262..627L} with an additional parameterisation to take into account the difference between the EPS theory and N-body simulations. The conditional mass function is expressed as
\begin{equation}
   f(M_1, z_1 | M_2, z_2) = G_0 \left( \frac{\sigma_1}{\sigma_2} \right)^{\gamma_1}  \left( \frac{\delta_2}{\sigma_2} \right)^{\gamma_2} f_\text{EPS}(M_1, z_1 | M_2, z_2),
\end{equation}
where $f_\text{EPS}(M_1, z_1 | M_2, z_2)$ is the conditional mass function given by the EPS theory. We denote $\sigma_1 = \sigma(M_1)$ and $\sigma_2 = \sigma(M_2)$, which are the mass variance of the matter density field linearly extrapolated to $z = 0$ and smoothed by a spherical tophat filter at $M_1$ and $M_2$. The density contrast is defined by $\delta_2 = 1.686/D(z_2)$, where $D(z)$ is the linear growth factor. The free parameters are $G_0$, $\gamma_1$ and $\gamma_2$. \cite{2008MNRAS.383..557P} calibrated these free parameters against the Millennium simulation \citep{2005Natur.435..629S} in the mass range between $10^{12} \hunit{\solarmass}$ and $10^{15} \hunit{\solarmass}$ and from $z = 0$ to $z = 4$. However, in this work, we are interested in growing halos at $z \geq 5$, and require that the mass resolution of the merger trees reaches the atomic cooling threshold ($\sim 10^7 - 10^8 \, h^{-1}M_\odot)$ in order to capture the majority of ionising sources during the epoch of reionisation. Therefore, we recalibrate the parameters against our simulations, which also accounts for updated cosmology.
\par
Following \cite{2008MNRAS.383..557P}, the cost function of the calibration is given by
\begin{equation}
    \mathcal{C}(G_0, \gamma_1, \gamma_2) = \sum \left[ \log_{10} f_\text{NS} -  \log_{10} f_\text{MC} \right]^2,
\end{equation}
where $f_\text{NS}$ and $f_\text{MC}$ are the conditional mass functions of the N-body and Monte Carlo merger trees respectively. We estimate $\log_{10} f_\text{NS}$ from L35N2650 and $\log_{10} f_\text{MC}$ using samples of 300 Monte Carlo merger trees for each descendant mass $M_2$. The fitting points calculated from the simulation are shown as black dots in Figure \ref{fig:fit}. We employ the particle swarm optimisation \citep{shi1998modified} to minimise the cost function. The best-fit parameters are accepted if they do not change for 100 iterations. Their values are given in Table \ref{table:params_mc}, and the fitting results are illustrated in Figure \ref{fig:fit}. Our best-fit parameters improve the cost function by $\Delta \mathcal{C} \approx -0.6$, compared with \cite{2008MNRAS.383..557P}. However, the best-fit result is still poor at $z = 5.5$ and $z = 10.1$. While a potential improvement is to employ weights for different mass or redshift ranges in the cost function, in Appendix \ref{sec:appendix}, we show that this approach cannot significantly improve the fitting results.

\subsubsection{Augmentation algorithm} \label{sec:augment}
The most important and difficult component of the augmentation is to decide whether a Monte Carlo tree is similar to an N-body tree. Instead of comparing entire trees,  \cite{2016ComAC...3....3B} decompose an N-body merger tree into many sub-branches, and match only one sub-branch every time with Monte Carlo realisations. A sub-branch is comprised of one descendant halo and all halos that directly merge into it. Hereafter, we refer to this structure as a "simple branch".
\par
We denote the mass of each progenitor in an N-body simple branch as $M_1$, $M_2$, ..., $M_n$ with $M_1 > M_2 > ... > M_n$, where $n$ is the number of the progenitors, and let $n_\text{cut}$ be the number of the progenitors whose mass is above a threshold $M_\text{cut}$. We use primed symbols for the same quantities of Monte Carlo trees. \cite{2016ComAC...3....3B} match N-body and Monte Carlo simple branches using:
\begin{itemize}
    \item[(a)] $n' \geq n_\text{cut}$, 
    \item[(b)] for $i=1,2,...,n_\text{cut}$, $|M_i - M'_i| < \xi M_i$,
    \item[(c)] for $i=n_\text{cut}+1,n_\text{cut}+2,...,n'$, $M'_i < M_\text{cut}$, 
\end{itemize}
where $\xi$ is a free parameter and controls the mass precision of the match. Once a match is found, N-body progenitors at $M_\text{halo} < M_\text{cut}$ are replaced by Monte Carlo halos in the same mass range. In the resulting hybrid structure, the descendant halo and progenitors with mass above $M_\text{cut}$ are from the original simple branch, while progenitors with mass below $M_\text{cut}$ are additional Monte Carlo halos from the match.
\par
In practice, relaxing the three matching criteria (a), (b) and (c) is necessary, since there is often no match even for large numbers of Monte Carlo realisations. \cite{2016ComAC...3....3B} increase $\xi$ by a factor of 1 + $\epsilon_\text{mass}$ after $N^\text{limit}_\text{mass}$ rejections. However, this only impacts the second condition. We have also found many cases where the first and third conditions are never satisfied. This problem was not reported in \cite{2016ComAC...3....3B}, and the reason might be that the mass range investigated in this work is much lower than in that paper. To address this issue, we increase $M_\text{cut}$ by a factor of $1 + \epsilon_\text{cut}$ after $N^\text{limit}_\text{cut}$ rejections. We do not allow $M_\text{cut}$ to be greater than either a maximum mass cut $M^\text{max}_\text{cut}$ or the mass of the most massive progenitor. Furthermore, a maximum number of trials $N^\text{limit}_\text{tot}$ is employed. Once this number of trials is reached, the algorithm is terminated and returns the input simple branch, with all progenitors below the minimum mass cut $M^\text{min}_\text{cut}$ removed. This treatment may remove some N-body halos without augmentation of Monte Carlo halos. However, in practice, we find that this situation occurs at a rate that is always smaller than $0.06 \%$ for a given snapshot.
\par
N-body merger trees have a special feature that should be taken into account in the comparison with Monte Carlo merger trees. When the halo finder fails to identify the descendant of an N-body halo in the next snapshot, it may try to search for the descendant in later snapshots. Hence, progenitors in an N-body simple branch are not always from the adjacent snapshot. However, this situation never happens for Monte Carlo merger trees. We follow \cite{2016ComAC...3....3B} to resolve the issue. In order to make the trees comparable, for a given N-body simple branch, we manually set all progenitors to be located in the previous snapshot relative to their descendant, and keep their mass unchanged (except for the most massive progenitor, whose mass is interpolated with time).
\par
N-body merger trees typically contain subhalos, which is an additional feature that Monte Carlo merger trees do not have. Following \cite{2016ComAC...3....3B}, we do not consider subhalos in the tree augmentation. Accordingly, we reconstruct a merger tree that only consists of host halos from an original N-body tree. The reconstruction proceeds forward with time. If the descendant of an N-body halo is a subhalo, we link it to the host of the subhalo. We neglect the descendant of a subhalo when building the host halo merger trees. We note that the reconstructed trees are only used during the Monte Carlo augmentation. When applying the augmented trees to semi-analytic models, the original links of N-body halos (including subhalos) are adopted. We note that these original links may be broken since some N-body halos are removed by the augmentation algorithm. Section \ref{sec:subhalo} will discuss the approach to fix the issue.  
\par
In reconstructed N-body merger trees, we have found many massive halos ($M_\text{halo} \gtrsim 10^{10} \hunit{\solarmass}$) that have no progenitors. In the original trees, these halos only have one subhalo progenitor whose host merges into a different target. When augmenting such halos, criteria (a) and (b) are automatically satisfied. However, we find that forcing criterion (c) overestimates the conditional mass function at $M_\text{halo} < M_\text{cut}$. Based on several experiments, we suggest the following modification, which can lead to more consistent conditional mass functions
\begin{itemize}
    \item[(c$'$)] if $n > 0$, for $i=n_\text{cut}+1,n_\text{cut}+2,...,n'$, $M'_i < M_\text{cut}$, otherwise  for $i=1,2,...,n'$, $M'_i < M^\text{max}_\text{cut}$.
\end{itemize}
\par
\begin{figure}
	\includegraphics[width=\columnwidth]{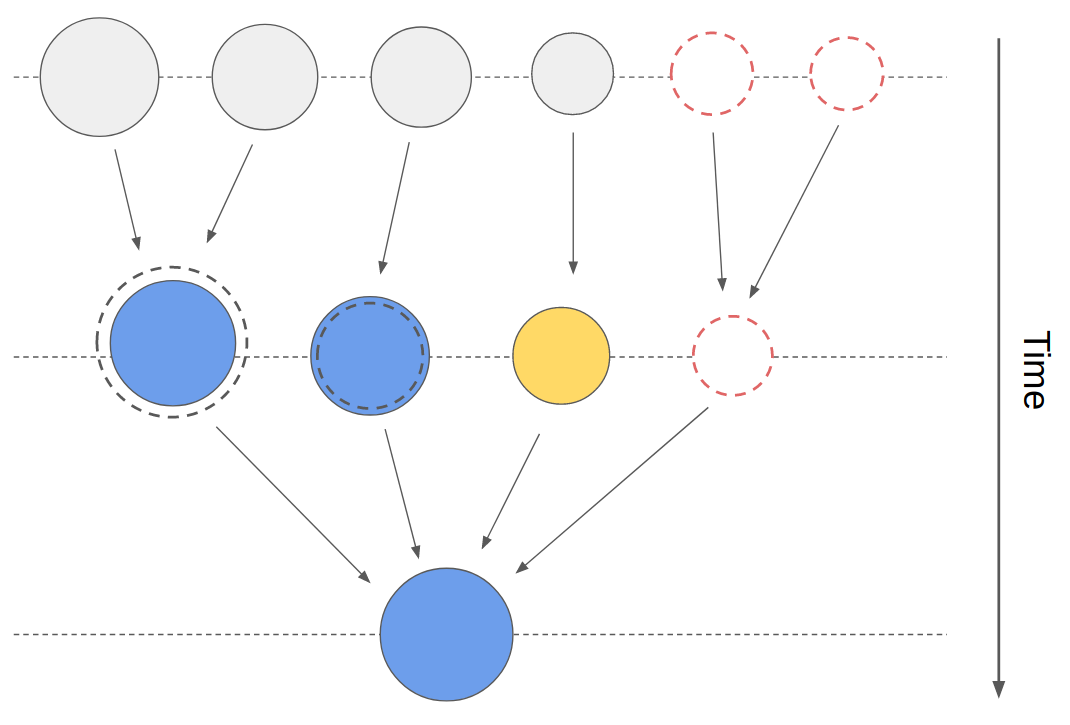}
    \caption{Schematic diagram of augmenting N-body halo merge trees. Solid and dashed circles represent N-body and Monte Carlo halos respectively, with radius proportional to halo mass. The blue and yellow circles form an N-body simple branch (defined in Section \ref{sec:augment}), which is compared with a Monte Carlo tree. Grey circles also represent N-body halos, but are not considered in this comparison. The algorithm removes halos with mass below $M_\text{cut}$, corresponding to the yellow circle. The progenitors of removed halos will not be taken into account in the next step. Red dashed circles represent Monte Carlo halos that are added to the N-body simulation. The Monte Carlo halos on the top are grown from its descendant using the \citet{2008MNRAS.383..557P} algorithm.}
    \label{fig:augment}
\end{figure}

Overall, given a simple branch in an N-body merger tree, our augmentation algorithm proceeds as follows:
\begin{enumerate}
    \item Set $N^\text{trial}_\text{cut} = 0$, $N^\text{trial}_\text{mass} = 0$, $N^\text{trial}_\text{tot} = 0$, $\xi = \xi_0$, $M_\text{cut} = M^\text{min}_\text{cut}$.
    \item Whenever a progenitor is at a non-adjacent snapshot of its descendant halo, put it to one previous snapshot of the descendant. If the progenitor is the most massive, interpolate its mass with time.
    \item \label{gmc} Generate a Monte Carlo simple branch using the same configuration as the given N-body branch. Increase $N^\text{trial}_\text{tot}$ by 1.
    \item Compare the N-body and Monte Carlo simple branches using criteria (a), (b) and (c$'$). If all three criteria are satisfied, go to step \ref{final}, otherwise, increase the corresponding counters:
        \begin{itemize}
            \item If criteria (a) or (c$'$) are false, increase $N^\text{trial}_\text{cut}$ by 1.
            \item If criterion (b) is false, increase $N^\text{trial}_\text{mass}$ by 1.
        \end{itemize}
    \item Relaxing the criteria when certain number of rejections is reached:
        \begin{itemize}
            \item If $N^\text{trial}_\text{cut} = N^\text{limit}_\text{cut}$, set $N^\text{trial}_\text{cut}=0$ and increase $M_\text{cut}$ by a factor of $1 + \epsilon_\text{cut}$. If $M_\text{cut}$ is greater than $M^\text{max}_\text{cut}$ or the mass of the most massive progenitors of the given simple branch, set it to be the minimum of these two values.
            \item If $N^\text{trial}_\text{mass} = N^\text{limit}_\text{mass}$, set $N^\text{trial}_\text{mass}=0$ and increase $\xi$ by a factor of $1 + \epsilon_\text{mass}$.
        \end{itemize} 
    \item Terminate the algorithm if $N^\text{trial}_\text{tot} = N^\text{limit}_\text{tot}$, otherwise go to step \ref{gmc}. 
    \item \label{final} Replace progenitors with mass below $M_\text{cut}$ at the N-body simple branch with Monte Carlo halos in the same mass range.
\end{enumerate}
We apply the augmentation algorithm to every halo in the N-body simulation backward with time, and grow new Monte Carlo halos using the \cite{2008MNRAS.383..557P} algorithm. A schematic diagram of the augmentation can be found in Figure \ref{fig:augment}.
\par
Free parameters in the algorithm are summarised in Table \ref{table:params_ag}. Ideally, if the conditional mass functions of Monte Carlo merger trees are consistent with the N-body simulations,
these parameters should primarily affect numerical efficiency and be insensitive to the results. However, as demonstrated in Figure \ref{fig:fit}, even with recalibrated parameters, the \cite{2008MNRAS.383..557P} algorithm is unable to reproduce all parts of the conditional mass functions of the N-body merger trees, particularly at the lower mass end and higher redshifts. For this reason, we find that the choice of the algorithm parameters impacts the resulting conditional mass functions. The values listed in Table \ref{table:params_ag} are chosen based on several experiments in order to obtain better consistency with the N-body simulations.
\par
To summarise, our augmentation algorithm builds on the method of \cite{2016ComAC...3....3B} by changing the mass cut $M_\text{cut}$ dynamically (and introducing the maximum mass cut $M^\text{max}_\text{cut}$). When applying the approach of \cite{2016ComAC...3....3B}, the result contains only Monte Carlo halos at $M_\text{halo} < M_\text{cut}$ and only N-body halos at $M_\text{halo} \geq M_\text{cut}$. In our approach, $M_\text{cut}$ is not a constant. The minimum and maximum mass cuts become the dividing lines of N-body and Monte Carlo halos. At the mass range in between, halo types are mixed. This modification averages the difference between the merger trees extracted from N-body simulations and those generated by the Monte Carlo algorithm.

\begin{figure}
	\includegraphics[width=\columnwidth]{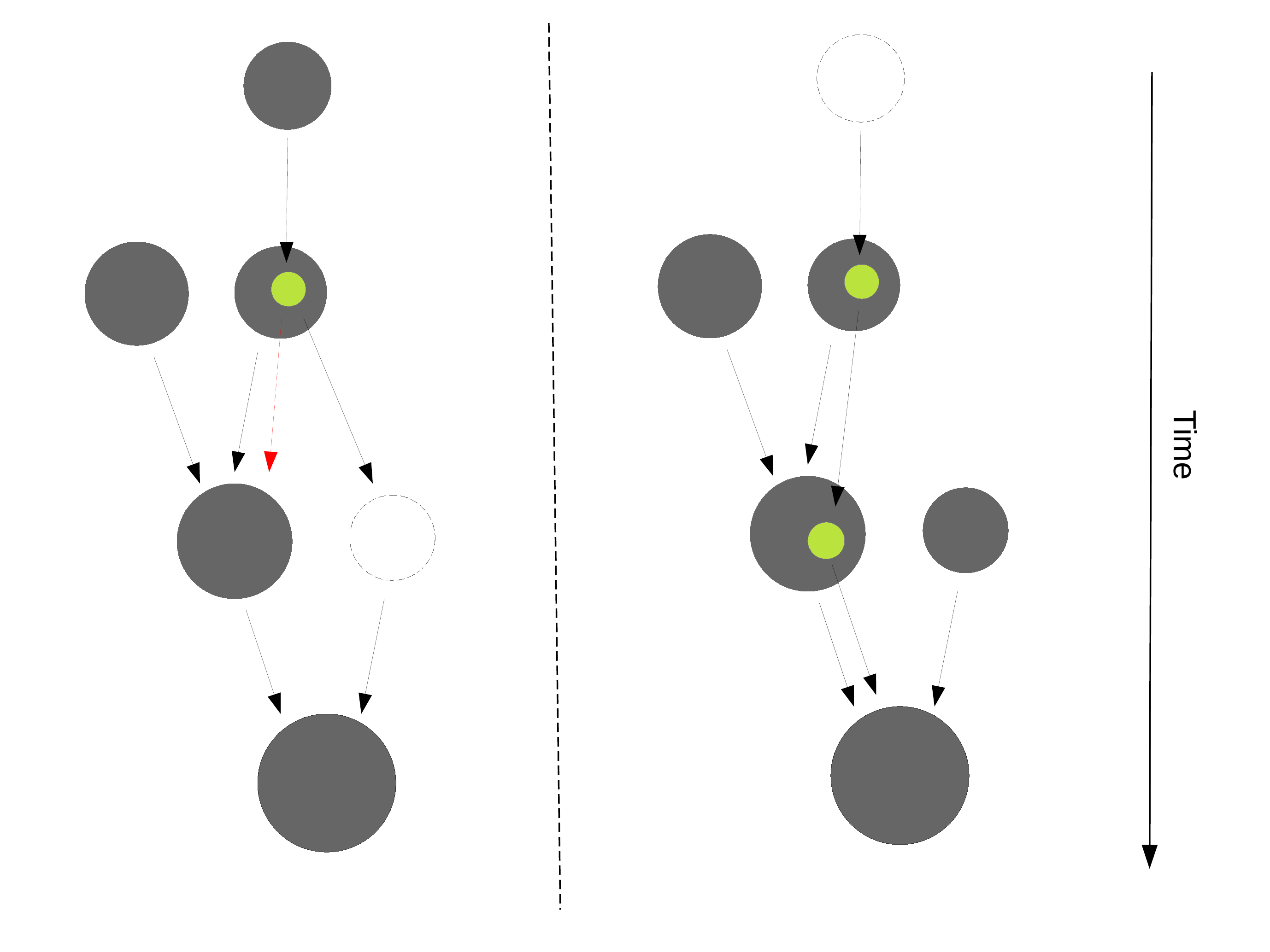}
    \caption{Schematic diagram of fixing subhalo trees. In both panels, black and green circles represent host halos and subhalos respectively. Empty circles correspond to a halo removed by the augmentation algorithm. In the left panel, a subhalo merges into a removed halo, and the host of the subhalo merges into a different target. We fix the problem by redirecting the merger target of the subhalo to the descendant of its host halo as shown by the red dashed arrow. In the right panel, a progenitor host halo of a subhalo is removed by the augmentation algorithm. Consequently, the whole corresponding subhalo tree becomes invalid. This issue can be fixed by preventing semi-analytic models from seeding a galaxy in a subhalo.}
    \label{fig:subhalo}
\end{figure}

\subsubsection{Fixing original subhalo trees} \label{sec:subhalo}
In N-body simulations, secondary progenitors may still be self-bound for a certain period  after a merger. Such objects are known as subhalos. During the tree augmentation, we reconstruct N-body merger trees that only include host halos. The reconstructed trees are only used in the comparison of Monte Carlo merger trees. In the application of the extended trees to semi-analytic models, we include subhalos from the original N-body trees. However, the augmentation algorithm removes an N-body halo if its mass is below $M_\text{cut}$, which may break an original subhalo tree. The left panel of Figure \ref{fig:subhalo} shows a such case, where a subhalo (green circle) merges into a removed halo (dashed circle), and the host of the subhalo merges into a different target. To fix the problem, we redirect the merger target of the subhalo to the descendant of its host halo as shown by the red dashed arrow. An additional case that is worth mentioning is illustrated in the right panel of Figure \ref{fig:subhalo}, where a progenitor host halo of a subhalo is removed during the Monte Carlo augmentation. Consequently, the whole corresponding subhalo tree should also be removed. An easier way to fix the issue is to prevent semi-analytic models from seeding a galaxy in such subhalos. This treatment is implemented in our application of the extended trees in Section \ref{sec:app}.

\subsubsection{Identifying the complete halo population} \label{sec:type}
A complete halo population cannot be obtained by applying the augmentation algorithm introduced in Section \ref{sec:augment}. The reason is that all Monte Carlo halos added by the algorithm will eventually merge into an N-body halo, while there are unresolved halos that do not interact with any N-body halo at the redshift range covered by the algorithm. This suggests that an additional catalogue of Monte Carlo halos are required to obtain a complete halo population.
\par
As a specific example in this work, we apply the augmentation algorithm at $z = 5$, adding Monte Carlo halos to the N-body merger trees backwards in time. However, at $z = 5$, the algorithm does not add new halos that are not resolved (between $M^\text{min}_\text{cut}$ and $M_\text{res}$). In addition, we also miss progenitors of such unresolved halos in earlier snapshots, resulting in an incomplete halo population. To fix this problem, we create an additional halo catalogue at $z = 5$, using masses and numbers drawn from the halo mass function of L35N2650. We use interpolation of a histogram instead of a fitting model for the halo mass function. We then generate trees for these halos using the \cite{2008MNRAS.383..557P} algorithm. Hereafter, Monte Carlo halos generated by the augmentation algorithm are labelled as MC-I, while those in the additional catalogue are referred to as MC-II.

\subsubsection{Applying to N-body simulations}
We apply the approach introduced in the proceeding sections to augment the N-body merger trees of L105N2048 from $z=5$ to $z=20$. We choose three levels of mass resolution: $M_\text{res} = 1.4 \times 10^8$, $5.7 \times 10^7$ and $3.2 \times 10^7 \hunit{\solarmass}$, corresponding to the atomic cooling threshold at $z = 5$, $z = 10$ and $z = 15$ respectively. These three extended halo catalogues are labelled as L105E5, L105E10 and L105E15. Their information is summarised in Table \ref{table:cat}.
\par
\begin{figure*}
	\includegraphics[width=0.97\textwidth]{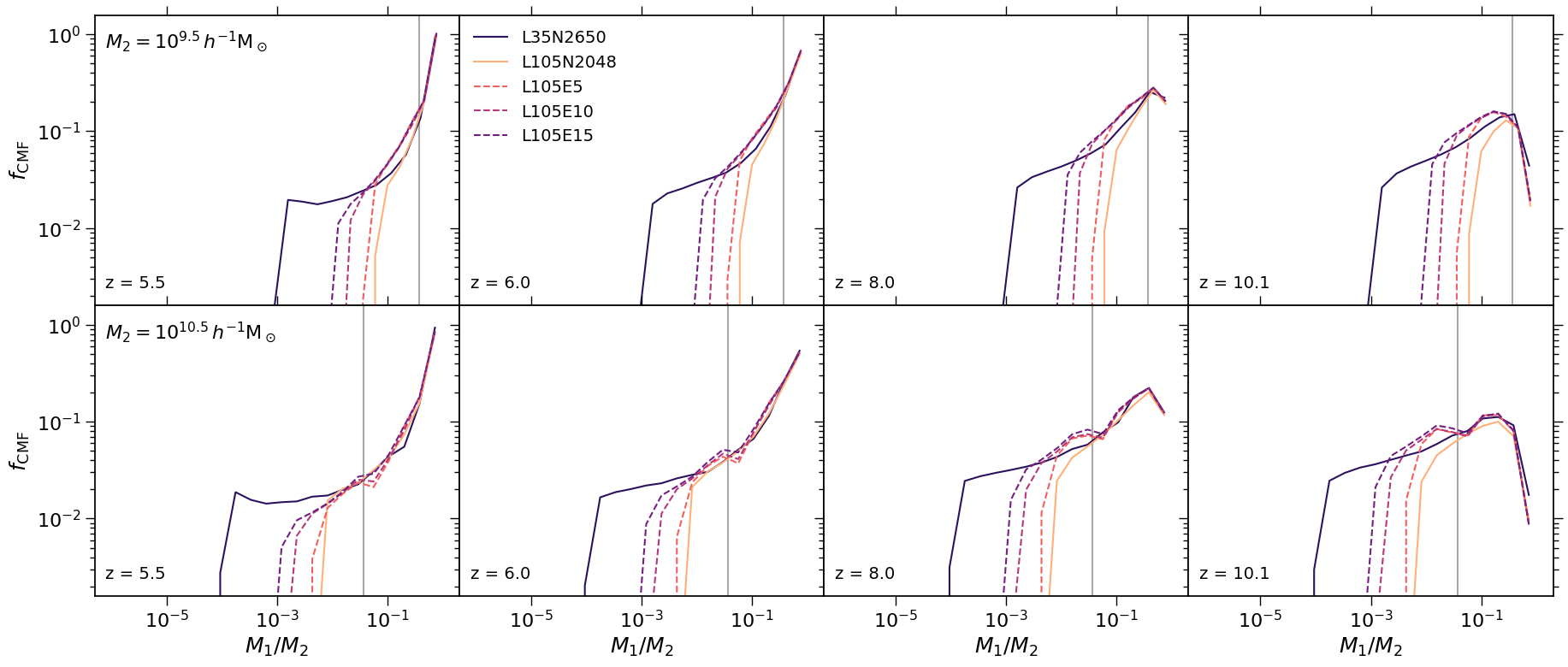}
    \caption{Comparisons of the conditional functions, defined by $d f_\text{CMF} / d \ln M_1$, of N-body and augmented merger trees. Solid lines are the results derived using L35N2650 and L105N2048. The information on these two N-body simulations can be found in Table \ref{table:cat}. Dashed lines are based on augmented halo merger trees, which are obtained by applying the algorithm described in Section \ref{sec:augment} to L105N2048. Darker colours correspond to higher mass resolution. The grey vertical lines show the minimum mass cut of the augmentation algorithm.}
    \label{fig:cmf_aug}
\end{figure*}

To test the results, we compare the conditional mass functions of augmented merger trees with our L35N2650 high resolution simulation in Figure \ref{fig:cmf_aug}. Upper and lower panels correspond to different descendant halo mass bins. The conditional mass functions (CMFs) of extended trees are shown as dashed lines, which broadly agree with L35N2650. Several discrepancies, e.g. the underestimation at the low mass end at $z = 5.5$, can be explained by the fact that the CMFs given by the \cite{2008MNRAS.383..557P} algorithm do not fully agree with the simulation as demonstrated in Figure \ref{fig:fit}. However, we find that this overestimation does not affect the stellar mass functions when applying a semi-analytic model to the augmented trees. We show this in Section \ref{sec:app}.
\par
\begin{figure*}
	\includegraphics[width=0.97\textwidth]{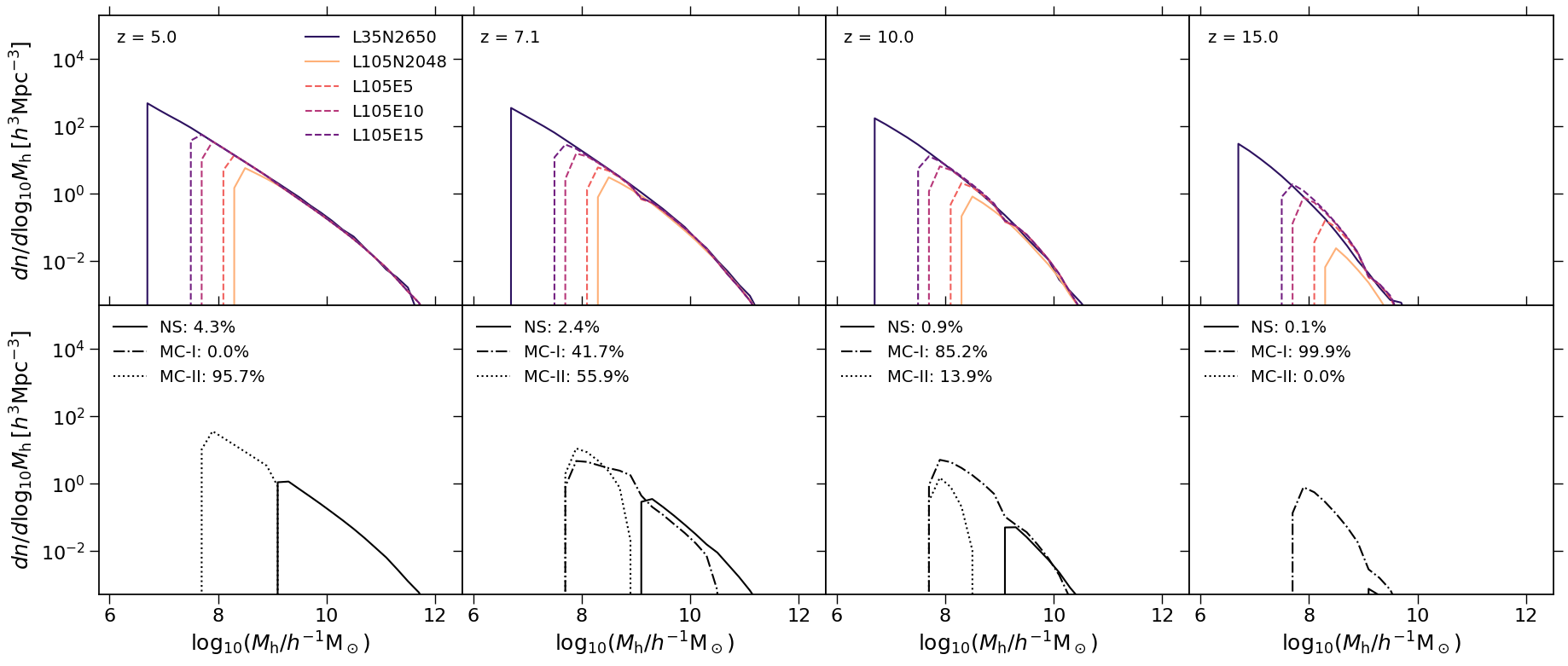}
    \caption{Upper panels: comparisons of the halo mass functions of N-body and extended halo catalogues. Solid lines are estimated from the N-body simulations, using L35N2650 and L105N2048. Their information can be found in Table \ref{table:cat}. Dashed lines are based on extended halo catalogues, which are obtained by applying the algorithm described in Section \ref{sec:algorithm} to L105N2048. The mass resolutions of L105E5, L105E10 and L105E15 correspond to the atomic cooling thresholds at $z$ = 5, 10 and 15 respectively. Darker colours correspond to higher mass resolution. Bottom panels: halo mass functions of N-body, MC-I and MC-II halos from L105E10. Their mass fractions are labelled in the top left corners. See Section \ref{sec:type} for the definition of MC-I and MC-II halos.}
    \label{fig:hmf}
\end{figure*}

The halo mass functions (HMFs) of the extended trees are demonstrated in the upper panels of Figure \ref{fig:hmf}. They show excellent agreement with L35N2650. The lower panels of the figure explicitly show the HMFs of N-body, MC-I and MC-II halos from L105E10. As defined in Section \ref{sec:type}, MC-I halos augment N-body merger trees, while MC-II halos are added to form a complete sample of halos, and are independent of N-body halos. While MC-II halos dominate the population at lower redshifts, MC-I halos are the main contributor at higher redshifts. Hence, both types of halos are necessary to calculate the halo abundance across all redshifts.  

\begin{table} 
    \centering
	\caption{Parameters of the tree augmentation algorithm.}
    \label{table:params_ag}
    \begin{tabular}{c|c}
	\hline 	\hline
    Symbol & Value \\
	\hline
    $\xi_0$ & 0.2 \\
    $\epsilon_\text{mass}$ & 0.2 \\
    $N^\text{limit}_\text{mass}$ & 50\\
    \hline
    $M^\text{min}_\text{cut}$ & $100 \, m_\text{p}$ $^\text{a}$\\
    $M^\text{max}_\text{cut}$ & $2500 \, m_\text{p}$ $^\text{a}$\\
    $\epsilon_\text{cut}$ & 2.0 \\
    $N^\text{limit}_\text{cut}$ & 5 \\
    \hline
    $N^\text{trial}_\text{tot}$ & 1000 \\
    \hline
    \end{tabular}
   \begin{tablenotes}
      \item $^\text{a}$ For L105N2048, $m_\text{p} = 1.17 \times 10^7 \hunit{\solarmass}$, which is the particle mass of the simulation. 
    \end{tablenotes}
\end{table}

\begin{table*}
	\caption{Information on halo catalogues used in this work.}
    \label{table:cat}
    \begin{tabular}{ccccccc}
    \hline \hline
    Name & Type & Box size $[h^{-1} \text{Mpc}]$ & Particle mass $[h^{-1} M_\odot]$ & Mass resolution $[h^{-1} M_\odot]$ \\ \hline
    L35N2650 & N-body simulation & 35 & $2.00 \times 10^5$ & - \\
    L105N2048 & N-body simulation & 105 & $1.17 \times 10^7$ & - \\ \hline
    L105E5 & Hybrid & 105 & - & $1.4 \times 10^8$ \\
    L105E10 & Hybrid & 105 & - & $5.7 \times 10^7$ \\
    L105E15 & Hybrid & 105 & - & $3.2 \times 10^7 $ 
        \\
    \hline
    \end{tabular}
    \begin{tablenotes}
         \item The mass resolutions of L105E5, L105E10 and L105E15 correspond to the atomic cooling threshold at $z$ = 5, 10 and 15 respectively.
    \end{tablenotes}
\end{table*}

\begin{figure*}
	\includegraphics[width=\textwidth]{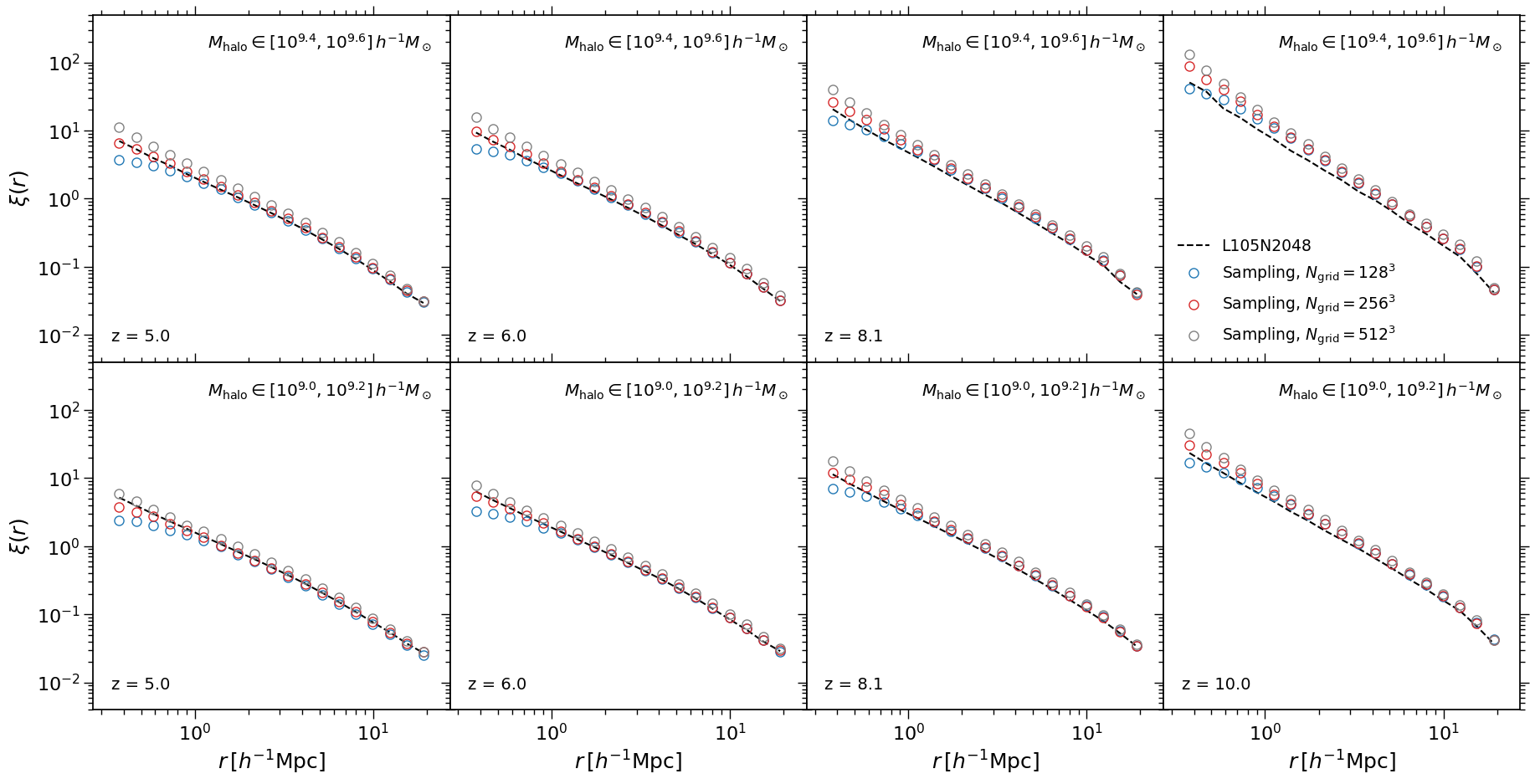}
    \caption{Comparison of two-point correlation functions produced using the random sampling method and estimated from N-body simulations. Empty circles are the results based on the random sampling method introduced in Section \ref{sec:bias}, with colours corresponding to different grid sizes as labelled on the top rightmost panel. Black dashed lines are estimated from the L105N2048 N-body simulations. Each row corresponds to a halo mass bin. These mass ranges are well resolved by L105N2048.}
    \label{fig:tpcf_s}
\end{figure*}

\begin{figure*}
	\includegraphics[width=\textwidth]{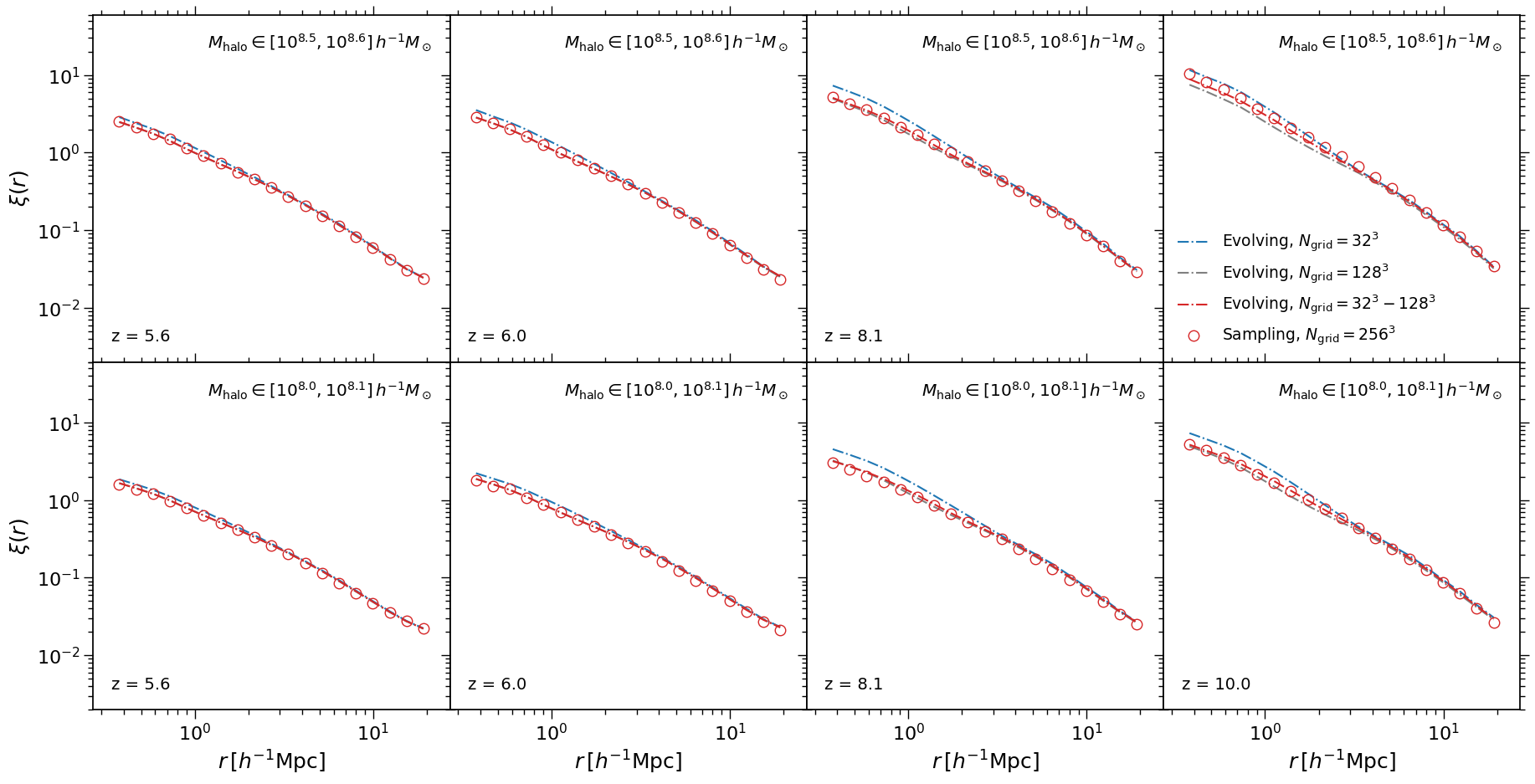}
    \caption{Comparison of two-point correlation functions produced using the evolving method and estimated from N-body simulations. Dash-dotted lines are the results based on the evolving method introduced in Section \ref{sec:evolve}. For blue and grey lines, grids with $32^3$ and $128^3$ cells are used to calculate the velocity field respectively, while for red lines, the adopted grid size varies with redshift, with $128^3$ cells at $z = 5 - 6$, $64^3$ cells at $z = 6 - 8$, and $32^3$ cells at $z > 8$. Red empty circles are the results obtained using the sampling method descibed in Section \ref{sec:bias}, which can be used to check the accuracy of the evolving method.}
    \label{fig:tpcf_e}
\end{figure*}

\begin{figure*}
	\includegraphics[width=\textwidth]{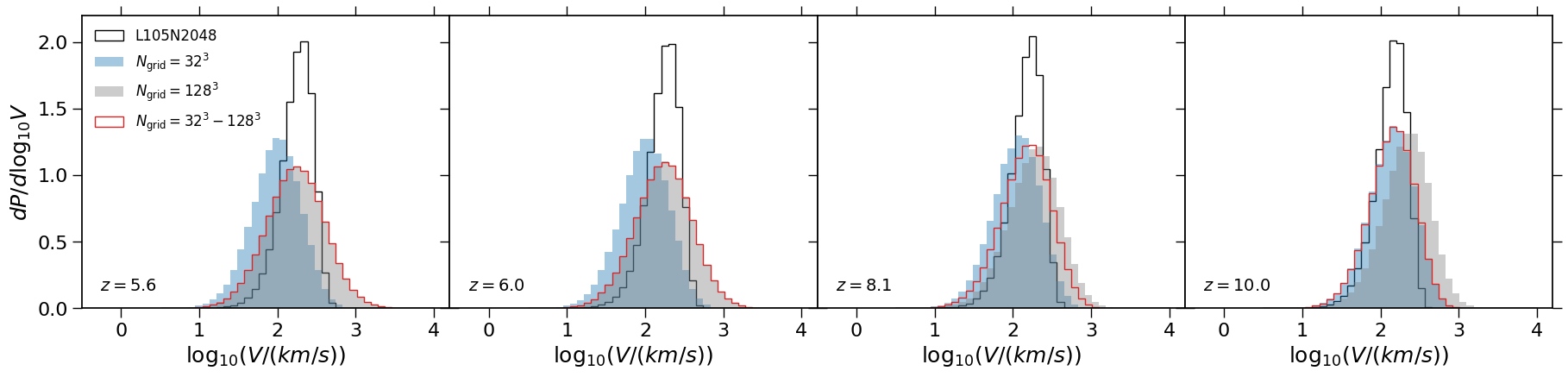}
    \caption{Peculiar velocity distributions of N-body and Monte Carlo halos. The velocities of Monte Carlo halos are derived using the method introduced in Section \ref{sec:evolve}. For blue and grey histograms, grids with $32^3$ and $128^3$ cells are used in the calculations, while for red histograms, the adopted grid size varies with redshift, with $128^3$ cells at $z = 5 - 6$, $64^3$ cells at $z = 6 - 8$, and $32^3$ cells at $z > 8$. The distributions of N-body halos are shown as black histograms.}
    \label{fig:vel}
\end{figure*}

\begin{figure*}
	\includegraphics[width=\textwidth]{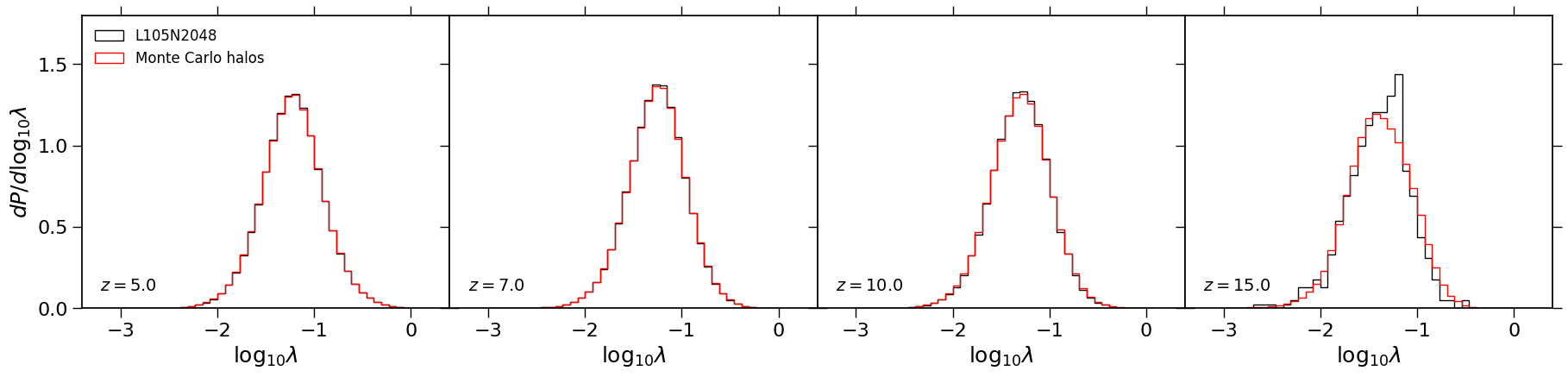}
    \caption{Spin distributions of N-body and Monte Carlo halos, plotted as back and red histograms respectively. The spin parameters of Monte Carlo halos are resampled from the distributions of N-body halos using a Gaussian kernel density estimator \citep[see e.g.][]{scott2015multivariate}. We only include N-body halos comprised of at least 100 particles to estimate their spin distributions, with subhalos excluded.}
    \label{fig:spin}
\end{figure*}

\subsection{Halo positions} \label{sec:pos}
When modelling reionisation, we require spatial information for halos within the extended halo catalogues. We aim to assign a position to every Monte Carlo halo and ensure that their two-point statistics agree with N-body simulations. Section \ref{sec:bias} discusses a random sampling method for placing MC-II halos within the simulation in the snapshot that the augmentation of the N-body merger trees is started, i.e. at $z = 5$. The method is then also used to verify our approach for evolving the position of Monte Carlo halos based on the position of their descendant, which is introduced in Section \ref{sec:evolve}.

\subsubsection{Populating halo positions} \label{sec:bias}
Monte Carlo halos can be populated into a simulation box using an analytic halo bias to transform the dark matter density field to a halo density field as a function of halo mass \citep{2013MNRAS.435..743D,2014MNRAS.442.3256A,2014MNRAS.441..646N,2015MNRAS.450.1486A,2020MNRAS.494.3294N}. In this work, the dark matter density field is estimated from the L105N2048 N-body simulation using the nearest grid point method. The result is represented as a cubic grid. To estimate the halo density field, we adopt the non-linear halo bias proposed by \cite{2015MNRAS.450.1486A}, which avoids negative density in underdense regions, and results in better two-point correlation functions on smaller scales. Halo positions are obtained by random sampling. We normalise the halo density field derived from the halo bias, and treat it as a one-dimensional discrete probability distribution. Then, at a given snapshot, we assign every Monte Carlo halo to a cell according to this probability and place it uniformly within the cell so that the number of halos in each cell follows the Poisson distribution. This approach does not depend on the normalisation of the halo density field and can be applied to any given number of Monte Carlo halos.
\par
To verify this method, we carry out a test within mass ranges that are well resolved by L105N2048. Specifically, we apply this method to $10^5$ samples, placing them within an empty box and measuring their two-point correlation functions. Then, we compare the results using N-body halos from L105N2048. We perform the test at $M_\text{halo} = 10^{9.1} \hunit{\solarmass}$ and $M_\text{halo} = 10^{9.5} \hunit{\solarmass}$ from $z = 5$ to $z = 10$ with different grid sizes. The results can be found in Figure \ref{fig:tpcf_s}, which shows good agreement with those estimated from the N-body simulation.
\par
The small scale clustering predicted by the random sampling method is affected by the choice of grid sizes. Halo positions within a cell of the grid are inaccurate since they are assumed to be uniformly distributed. As expected, the two-point correlations obtained using a $128^3$ grid (shown as blue circles in Figure \ref{fig:tpcf_s}) are underestimated at separations smaller than $0.8 \hunit{\rm Mpc}$, which is equal to the cell size of the grid. In terms of the results using a $512^3$ grid (grey circles), they have slightly larger clustering amplitudes over all scales than those using a $256^3$ grid (red circles). A potential reason could be that the estimation of the dark matter density field becomes noisy when a larger number of cells are used. For the following applications, we adopt a $256^3$ grid for the random sampling method. This choice is appropriate since the corresponding cell size ($0.4 \hunit{\rm Mpc}$) is smaller than the characteristic size of ionising regions \citep[e.g.][]{2006MNRAS.365..115F}.
\par
Unfortunately, we are unable to do the same test for Monte Carlo halos in the extended halo catalogues. This is because a complete sample of N-body halos at these mass ranges is only available in L35N2650, for which the box size is not sufficient to estimate two-point statistics. However, we note that the linearity of halo density fields increase towards lower halo mass, implying that the results are likely to be improved at $M_\text{halo} \lesssim 10^{8} \hunit{\solarmass}$. This argument indicates that the results in Figure \ref{fig:tpcf_s} are conservative for estimating the accuracy of the method. Hence, our method can be safely applied to the mass ranges that we are interested in.

\subsubsection{Evolving halo positions} \label{sec:evolve}
Evolution in the clustering of halos is influenced by their peculiar motions. Our approach of evolving halo positions is based on the linear continuity equation. We again divide the L105N2048 box into a cubic grid with $256^3$ cells. For Monte Carlo trees at $t_1$, the first step is to place the halos into the same cell as their direct descendant at $t_2$. We assume that the spatial distribution of the halos at $t_1$ can be described by a halo density field denoted as $\mathcal{D} (\vec{x}, t_1)$. The idea is to move these halos using a velocity field such that their spatial distribution becomes a desired halo density field denoted as $\mathcal{D}  (\vec{x}, t_2)$. We assume that this process can be described by the linear continuity equation. If $\Delta t = t_2 - t_1$ is small, the velocity field can be obtained by
\begin{equation}
    \nabla \vec{v}(\vec{x}, t_2) = - \frac{1}{\Delta t} \left[ \mathcal{D} (\vec{x}, t_1) - \mathcal{D}  (\vec{x}, t_2) \right]
\end{equation}
In the linear regime, we want 
\begin{equation}
    \mathcal{D} (\vec{x}, t_1) = b(M_1, t_1) \delta_\text{DM} (\vec{x}, t_1)
\end{equation}
where $M_1$ is the mass of the Monte Carlo halos and $b(M, t)$ is the linear halo bias. After a forward evolution, the change of the density field for halos at $t_1$ with mass $M_1$ is contributed from both the variation of the background dark matter density field and local interactions such as smooth mass accretion and mergers. Although a detailed model that considers all the effects is complicated, we find that evolving halo positions using the following expression for $\mathcal{D} (\vec{x}, t_2)$ can lead to reasonable two-point statistics.
\begin{equation}
    \mathcal{D} (\vec{x}, t_2) = b(M_1/\bar{\mu}_\text{R}, t_2) \delta_\text{DM} (\vec{x}, t_2),
\end{equation}
where $\bar{\mu}_\text{R}$ is the mean mass ratio between the progenitor and descendant halos.
\par
Then, it is straightforward to compute the velocity field using the Fourier transform. The velocity field in $k$-space can be written as 
\begin{equation} \label{eqn:vel}
    \vec{v}(\vec{k}, t_2) = b(M_1/\bar{\mu}_\text{R}, t_2) \vec{u}(\vec{k}, t_2) - b(M_1, t_1) \vec{u}(\vec{k}, t_1)
\end{equation}
with
\begin{equation}
    \vec{u}(\vec{k}, t) = \frac{i \vec{k}}{\Delta t k^2}  \delta_\text{DM} (\vec{k}, t),
\end{equation}
The real space velocity field then can be obtained using the inverse Fourier transform. Since $\vec{u}(\vec{k}, t)$ is independent of halo mass, we only need to perform the Fourier transform once per snapshot, and the velocity can be calculated per halo, without any mass bins. This advantage is only available when the halo bias and the dark matter density field are separable. For the linear halo bias, we adopt the fitting model given by \cite{2010ApJ...724..878T}.
\par
We apply this method to all extended halo catalogues and find that the choice of grid sizes to calculate $\vec{u}(\vec{k}, t)$ can affect the results. In Figure \ref{fig:vel}, we show that the median velocity of Monte Carlo halos is underestimated at $z \sim 5$ using a $32^3$ grid and is overestimated at $z \sim 10$ using a $128^3$ grid. This trend is expected. The density field should not be over smoothed, as this loses the information on density peaks. On the other hand, the halo bias increases rapidly with redshift, in which case the halo density field cannot be described by the linear bias. Smoothing the density field over larger regions can increase the linearity.
\par
To verify the two-point correlation functions predicted by the evolving method, we have to use the sampling method introduced in the previous section. A direct comparison with L35N2650 is not feasible due to its limited box size, and the accuracy of this indirect approach is confirmed in the previous section. Figure \ref{fig:tpcf_e} compares the two-point correlation functions obtained using the sampling and evolving methods. Since halo positions are evolved backwards with time, when a $32^3$ grid is used, the errors due to the underestimation of the halo velocity accumulate towards higher redshifts, which results in the overestimation of the two-point correlation functions at $z \gtrsim 6$ (see blue dash-dotted lines). Overall, we find good agreement between the results based on both methods, particularly on large scales.
\par
Based on the discussion above, we have decided to vary the grid size with redshift when evolving halo positions. Specifically, we use a $128^3$ grid at $z = 5 - 6$, a $64^3$ grid at $z = 6 - 8$, and a $32^3$ grid at $z > 8$. This treatment results in both consistent two-point correlation functions and velocity distributions, which are shown as red dash-dotted lines and red histograms in Figure \ref{fig:tpcf_e} and Figure \ref{fig:vel} respectively.

\subsection{Spin parameters} \label{sec:spin}
Many semi-analytic models use the halo spin parameter (defined by \cite{2001ApJ...555..240B}) to compute quantities  including disk size and star formation rate. To facilitate this, we sample the spin parameter of Monte Carlo halos using the spin distributions estimated from the N-body simulation. At $ z \geq 5$, negligible dependence on halo mass is found in the spin distributions of our simulations, which is consistent with \cite{2008ApJ...678..621K} and \cite{2016MNRAS.459.2106A}. The mass independent spin distributions can be described by a log-normal distribution \citep[e.g.][]{1998ApJ...507..601V,2008ApJ...678..621K} or a modified profile taking into account the long tail of low spins \citep[e.g.][]{2007MNRAS.376..215B,2016MNRAS.459.2106A}. In this work, we adopt a non-parametric approach. We train a Gaussian kernel density estimator \citep[see e.g.][]{scott2015multivariate} using samples from our N-body simulations (in $\log_{10} \lambda$ space), and assign the spin of Monte Carlo halos by resampling from the density estimator. We choose the bandwidth of the density estimator according to Scott’s Rule \citep{scott2015multivariate}.
\par
In Figure \ref{fig:spin}, black and red histograms are the spin distributions based on N-body and Monte Carlo halos respectively. When assembling N-body halos to estimate the spin distributions, we only include halos comprised of at least 100 particles and exclude all subhalos. Our results illustrate excellent agreement between the resampled and original distributions by construction. We note that our approach can be generalised to the case where spin parameter is tightly correlated with halo mass by splitting the total sample into several mass bins and applying the kernel density estimator to each subsample.

\begin{figure*}
	\includegraphics[width=\textwidth]{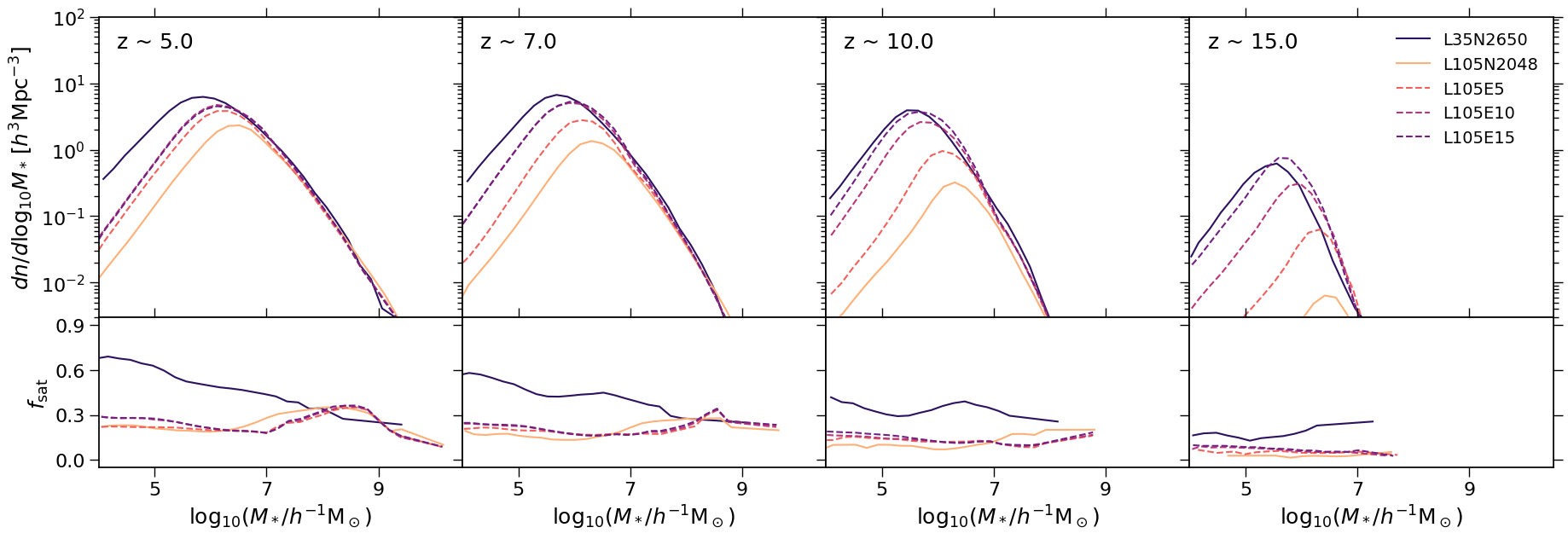}
    \caption{Upper panels: stellar mass functions predicted by the \textsc{meraxes} semi-analytic model. Lower panels: satellite fractions as a function of stellar mass. For all panels, solid lines use the original halo merger trees from our N-body simulations. Dashed lines are the results based on extended catalogues, which consist of both N-body and Monte Carlo halos. Darker colours correspond to higher mass resolution. The information on each halo catalogue as labelled in the top right corner can be found in Table \ref{table:cat}.}
    \label{fig:smf}
\end{figure*}

\begin{figure*}
	\includegraphics[width=\textwidth]{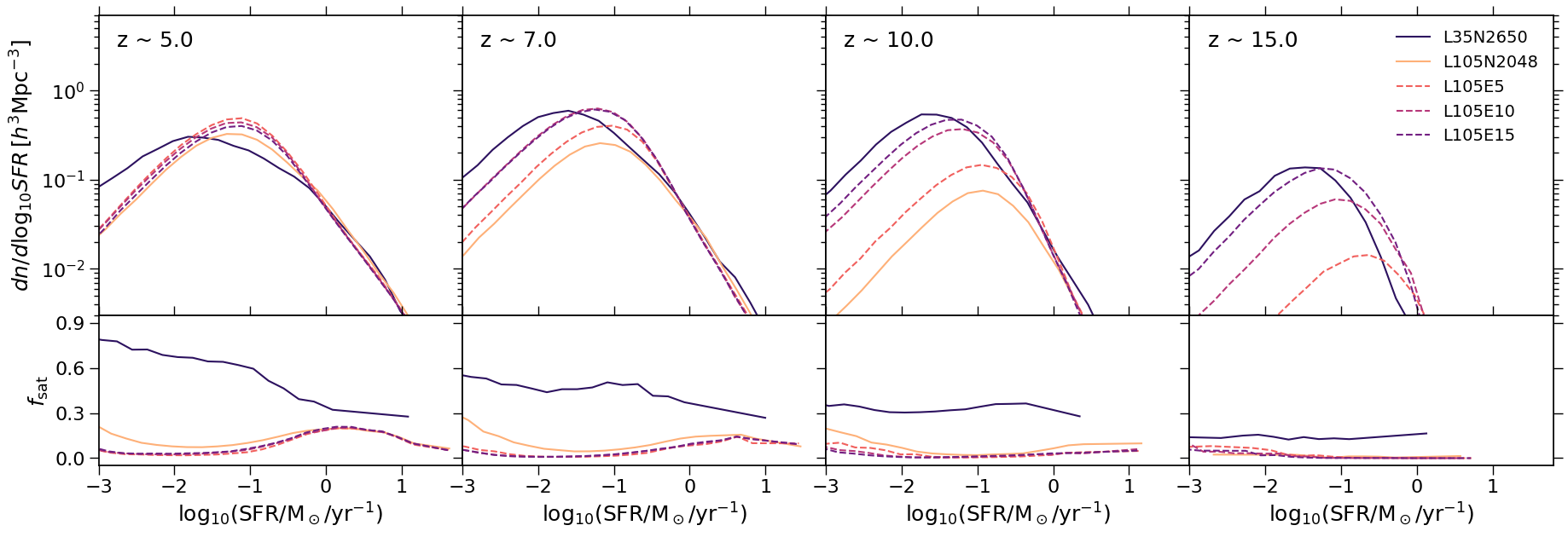}
    \caption{Upper panels: star formation rate functions predicted by the \textsc{meraxes} semi-analytic model. Lower panels: satellite fractions as a function of star formation rate. For all panels, solid lines use the original halo merger trees from our N-body simulations. Dashed lines are the results based on extended catalogues, which consist of both N-body and Monte Carlo halos. Darker colours correspond to higher mass resolution. The mass resolutions of L105E5, L105E10 and L105E15 are the atomic cooling thresholds at $z$ = 5, 10 and 15 respectively. The information on each halo catalogue as labelled in the top right corner can be found in Table \ref{table:cat}.}
    \label{fig:sfrf}
\end{figure*}

\section{Application to Meraxes} \label{sec:app}
We apply both the N-body and extended halo catalogues to the \textsc{meraxes} semi-analytic model \citep{2016MNRAS.462..250M}. In addition to the implementation of several key galaxy formation processes including radiative cooling, star formation and supernova feedback, the \textsc{meraxes} model is coupled with \textsc{21cmfast} \citep{2007ApJ...669..663M} to realise inhomogeneous reionisation feedback and to predict reionisation related properties such as the global neutral fraction and 21cm power spectra. The \textsc{meraxes} model only seeds galaxies in halos whose mass is above the atomic cooling threshold. We adopt the same parameters as \cite{2016MNRAS.462..250M} but note that the model predictions can be different from \cite{2016MNRAS.462..250M} due to the use of different halo merger trees. However, the main focus of this work is to demonstrate the consistency between the N-body and extended halo catalogues and to illustrate the consequences of adopting different halo mass resolutions rather than to present a model that satisfies all current observational constraints.

\begin{figure}
	\includegraphics[width=\columnwidth]{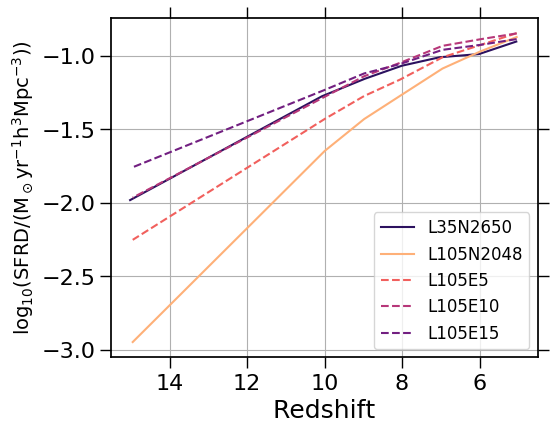}
    \caption{Star formation rate density predicted by the \textsc{meraxes} semi-analytic model. Solid lines use the original halo merger trees from our N-body simulations. Dashed lines are the results based on extended catalogues, which consist of both N-body and Monte Carlo halos. Darker colour corresponds to higher mass resolution. The mass resolutions of L105E5, L105E10 and L105E15 are the atomic cooling thresholds at $z$ = 5, 10 and 15 respectively. The information on each halo catalogue as labelled in the bottom right corner can be found in Table \ref{table:cat}.}
    \label{fig:sfrd}
\end{figure}

\begin{figure}
	\includegraphics[width=0.95\columnwidth]{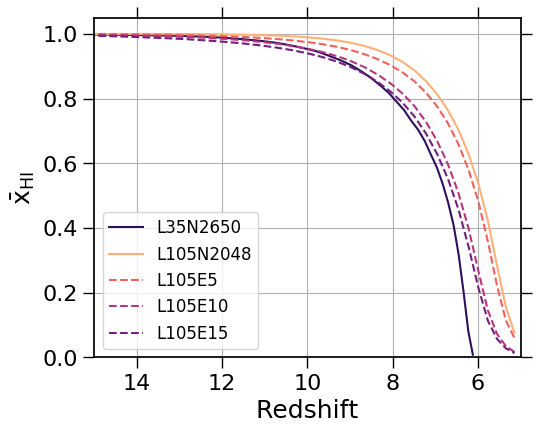}
    \caption{Volume-weighted neutral fractions predicted by the \textsc{meraxes} model. Solid lines and dashed lines are the results based on N-body and extended halo catalogues respectively. Darker colours correspond to higher mass resolution. The mass resolution of L105E5, L105E10 and L105E15 are the atomic cooling thresholds at $z$ = 5, 10 and 15 respectively. See Table \ref{table:cat} for the information on these halo catalogues.}
    \label{fig:xH}
\end{figure}

\begin{figure*}
	\includegraphics[width=\textwidth]{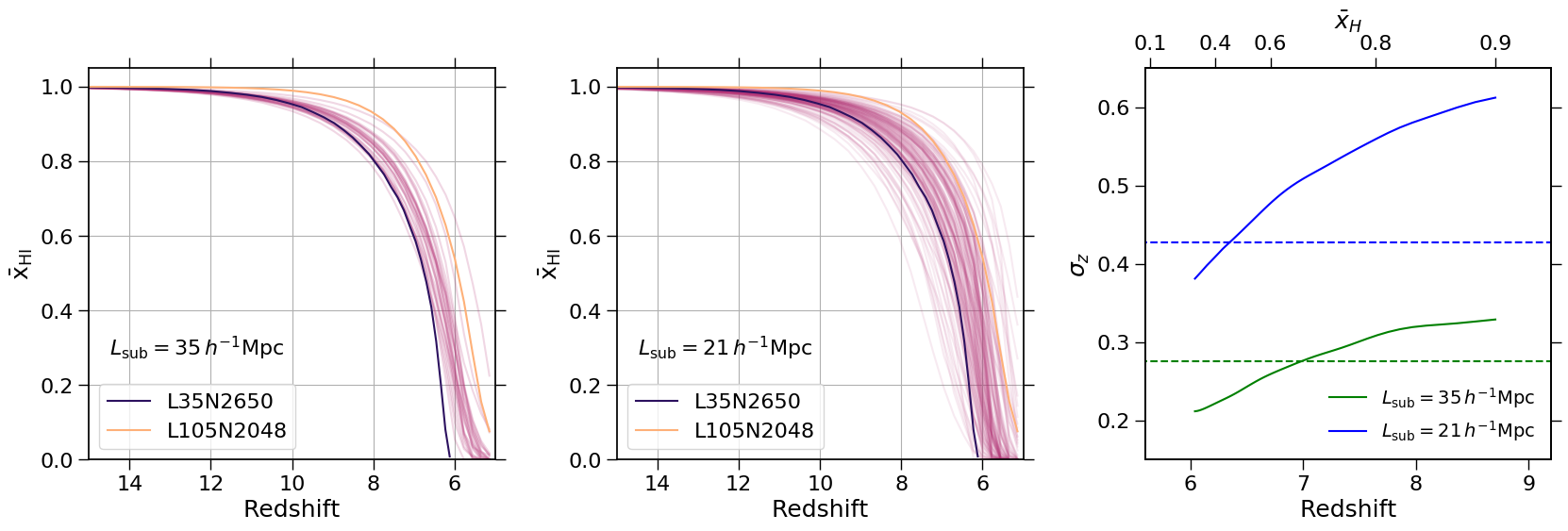}
    \caption{Left and middle panels show the reionisation histories in subvolumes with side lengths of $35 \hunit{\rm Mpc}$ and $21 \hunit{\rm Mpc}$ respectively. The latter is roughly equal to the maximum bubble size that we choose for \textsc{21cmfast}. These results are based on L105E10. In the right panel, solid lines show the standard deviations of redshift in subvolumes at fixed neutral fractions. Redshifts on the bottom axis are converted using the mean relation of the entire volume. The deviations are compared with the analytic predictions of \citet{2004ApJ...609..474B}, which are shown as dashed lines.}
    \label{fig:xH_sub}
\end{figure*}

\subsection{Galaxy properties} \label{sec:gal}
Figures \ref{fig:smf}, \ref{fig:sfrf} and \ref{fig:sfrd} demonstrate the stellar mass functions, star formation rate functions, and star formation rate densities predicted by \textsc{meraxes} respectively.
L35N2650 is a small volume N-body simulation with very high mass resolution, which is used to verify the results based on the extended halo catalogues. The predicted galaxy properties using L35N2650 and extended trees are shown as purple solid and dashed lines respectively. We find a difference in the peaks of both the stellar mass and star formation rate functions, which may result from the fact that Monte Carlo merger trees do not contain subhalos. This point is illustrated in the lower panels of Figures \ref{fig:smf} and \ref{fig:sfrf}, where we show that L35N2650 provides significantly higher satellite fractions than the extended halo catalogues, particularly at the low stellar mass and low star formation rate ends. In \textsc{meraxes}, all gas infalling into a friends-of-friends group is assumed to be accreted onto the central galaxy. Therefore, satellite galaxies have less fuel to form stars. Despite this disagreement, we find excellent agreement between the cosmic star formation rate densities obtained using L35N2650, L105E10 and L105E15 at $z < 10$. The result based on L105E15 shows higher star formation rate density than L35N2650 at $z > 10$. However, L35N2650 has a higher mass resolution. This is likely due to the overestimation of the halo mass functions at these redshifts as illustrated in Figure \ref{fig:hmf}.
\par
An additional finding is that the effect of mass resolution does not seem to be cumulative. While the mass resolutions of L105E5, L105E10 and L105E15 are different (and all above the atomic cooling threshold at $z = 5$), in Figure \ref{fig:smf}, their corresponding stellar mass functions overlap at $z = 5$. Figure \ref{fig:sfrd} also shows that the star formation rate densities predicted by the extended trees converge towards $z = 5$. These findings are non-trivial. We note that even if a halo is below the atomic cooling threshold at a given redshift, it can still host a galaxy. The reason is that the atomic cooling threshold increases with redshift, and as long as any progenitor of a halo is above the cooling limit, the halo will contain a galaxy. Therefore, we should not expect that halo catalogues with different mass resolution produce similar stellar mass and star formation rate functions towards $z = 5$. On the contrary, our results indicate that if all halos above the atomic cooling threshold at a given redshift are resolved, an ability to resolve less massive halos at an earlier time has little effect on predicted galaxy properties such as the stellar mass and star formation rate functions at the given redshift.

\subsection{Reionisation histories}
Having demonstrated the galaxy properties based on both N-body and extended halo catalogues, we now focus on the predictions of cosmic reionisation. The end of reionisation is known to be too rapid in simulations that do not resolve all faint galaxies or do not have a sufficiently large volume \citep{2004ApJ...609..474B,2014MNRAS.439..725I}. We therefore expect that the predictions of the reionisation history are sensitive to both halo mass resolution and simulation volume. 
\par
Figure \ref{fig:xH} illustrates the effect of halo mass resolution on the predicted volume-weighted neutral fractions. We see a difference between results using direct N-body merger trees (from L35N2650 and L105N2048). However, it is not straightforward to interpret this due to the different simulation volumes. Our extended halo catalogues (L105E10 and L105E15) have the same volume as L105N2048 and produce consistent star formation rate densities with L35N2650. Figure \ref{fig:xH} shows that the end of reionisation occurs earlier in L105E10 and L105E15 than in L105N2048, which confirms that mass resolution has an impact on the reionisation history. This is expected since reionisation is sensitive to cumulative star formation. A similar result was previously obtained by \cite{2018MNRAS.480.2628F}. We note that our results only indicate a minimum requirement of the mass resolution for predicting convergent reionisation histories, since our model neglects star formation below the atomic cooling threshold, which may also provide a non-negligible contribution to reionisation \cite[e.g.][]{2014MNRAS.442.2560W}.
\par
Small box simulations are known to suffer from both cosmic variance and lack of large scale modes \citep[e.g.][]{2004ApJ...609..474B}. We demonstrate this effect using subvolumes of the L105E10 extended halo catalogue. In the left and middle panels of Figure \ref{fig:xH_sub}, we show reionisation histories in two different sizes of subvolumes, having $L_\text{sub} = 35 \hunit{\rm Mpc}$ and $L_\text{sub} = 21 \hunit{\rm Mpc}$. The former has the same volume as L35N2650, while the latter is roughly equal to the maximum bubble size that we choose in the \textsc{21cmfast} algorithm within \textsc{meraxes}. Each subvolume contains different amounts of large scale power, leading to a rapid end of reionisation in each case, but at a range of redshifts. This explains the deviation of the shape of the late time reionisation history in L35N2650 from the predictions based on L105E10 and L105E15. The large volume simulations average cosmic variance shown within subvolumes in Figure \ref{fig:xH_sub}.
\par
In the right panel of Figure \ref{fig:xH_sub}, we compare the standard deviation of redshift at fixed neutral fractions in the subvolumes (solid lines) with the analytic prediction of \cite{2004ApJ...609..474B} (dashed lines). They pointed out that the difference of the collapse fraction in random regions of the Universe can be interpreted as an offset in redshift with respect to the cosmic mean. The scatter of the offset can be calculated from the critical collapse fraction, and be related to the width or duration of the reionisation history by equating it to the size of a particular reionisation region \citep{2004Natur.432..194W}. Despite the complexities in \textsc{meraxes}, the analytic prediction provides a reasonable estimation of cosmic variance. Overall, our results reinforce the importance of a large volume for cosmic reionisation simulations, which has also been highlighted by previous studies \citep[e.g.][]{2006MNRAS.369.1625I,2014MNRAS.439..725I,2020arXiv200406709D}.
\par
In addition, our results show that resolving all halos above the atomic cooling threshold across whole cosmic reionisation is important for calculating a converged reionisation history. \cite{2015ApJ...802L..19R} analysed the joint observational constraints of Thomson scattering optical depth measured by \cite{2016A&A...594A..13P} and cosmic star formation rate density estimated by \cite{2014ARA&A..52..415M}, suggesting that cosmic reionisation happens at $6 \lesssim z \lesssim 10$. Our results imply that simulations should reach at least the atomic cooling threshold at $z = 10$ in order to explore such reionisation scenarios. The decrease of the atomic cooling threshold with increasing redshift places constraints on the required halo mass resolution of simulations towards the beginning of reionisation.

\section{Summary} \label{sec:summary}
In this paper, we present a hybrid method to compute high resolution halo merger trees within large volume N-body simulations for semi-analytic reionisation models, which is based on the work of \cite{2016ComAC...3....3B}. As an application, we extend the mass resolution of halo merger trees extracted from the Genesis N-body $105 \, h^{-1}\text{Mpc}$ simulation box at $z \geq 5$. We verify the results using a small N-body simulation with very high resolution, and find good agreement for the halo mass functions. We also introduce a method to assign and evolve the position of Monte Carlo halos. The resulting two-point correlation functions are consistent with N-body simulations at separations greater than $0.4 \hunit{\rm Mpc}$. In the application to the \textsc{meraxes} semi-analytic model, the extended halo catalogues provide significant improvements on the predicted galaxy properties and reionisation history.
\begin{itemize}
    \item The decreasing atomic cooling threshold requires simulations to have higher mass resolution towards higher redshifts. Our model confirms that the faint sources at the beginning of reionisation can have a significant impact on the reionisation history, and therefore
    resolving the atomic cooling threshold throughout reionisation is necessary for reliable calculations of the reionisation history.
    \item The end of reionisation is predicted to be too rapid in simulations that either fail to resolve all faint galaxies or have a too small volume, putting demands on halo mass resolution and simulation volume. Using our extended tree algorithm, we show that the convergent predictions of the late stage reionisation history need both large volumes ($L_\text{box} \gtrsim 100 \hunit{\rm Mpc}$) and resolution of the atomic cooling threshold across the whole reionisation history.
    \item If all halos above the atomic cooling threshold at a given redshift are resolved, resolving even smaller halos at higher redshifts has negligible effect on predictions of galaxy population properties from the \textsc{meraxes} semi-analytic model at the given redshift.
\end{itemize}
\par
Our methodology provides a powerful tool to achieve desired mass resolution in large volumes. The largest extended halo catalogue obtained in this work has the mass resolution at $M_\text{halo} = 3.2 \times 10^7 \hunit{\solarmass}$ in a $105 \hunit{\rm Mpc}$ box, equivalent to an N-body simulations with $\sim 6800^3$ particles. Given the efficiency of the Monte Carlo algorithms, our approach can be applied to larger volumes (several hundred Mpc on each side), which are necessary for studying the statistics of reionisation including X-ray heating and global 21cm signal during cosmic dawn.

\section*{Acknowledgements}
We thank the anonymous referee for providing a detail report to improve the quality of the paper. This research was supported by the Australian Research Council Centre of Excellence for All Sky Astrophysics in 3 Dimensions (ASTRO 3D), through project number CE170100013. This work was performed on the OzSTAR national facility at Swinburne University of Technology. OzSTAR is funded by Swinburne University of Technology and the National Collaborative Research Infrastructure Strategy (NCRIS). YQ thanks Yuxiang Qin and Bradley Greig for useful discussions.
\par
We acknowledge the use of the following software: \textsc{astropy} \footnote{http://www.astropy.org} \citep{astropy:2013, astropy:2018}, \textsc{corrfunc} \citep{10.1007/978-981-13-7729-7_1,2020MNRAS.491.3022S}, \textsc{cython} \citep{2011CSE....13b..31B}, \textsc{hmf} \citep{2013A&C.....3...23M}, \textsc{ipython} \citep{2007CSE.....9c..21P}, \textsc{matplotlib} \citep{2007CSE.....9...90H}, \textsc{numpy} \citep{2011CSE....13b..22V}, \textsc{pandas} \citep{mckinney-proc-scipy-2010}, \textsc{seaborn}\footnote{https://github.com/mwaskom/seaborn} and \textsc{scipy} \citep{scipy}.

\section*{Data availability}
The data underlying this article will be shared on reasonable request to the corresponding author.

\bibliographystyle{mnras}
\bibliography{references}

\appendix
\section{Additional calibrations of the Parkinson algorithm} \label{sec:appendix}
\begin{figure*}
	\includegraphics[width=\textwidth]{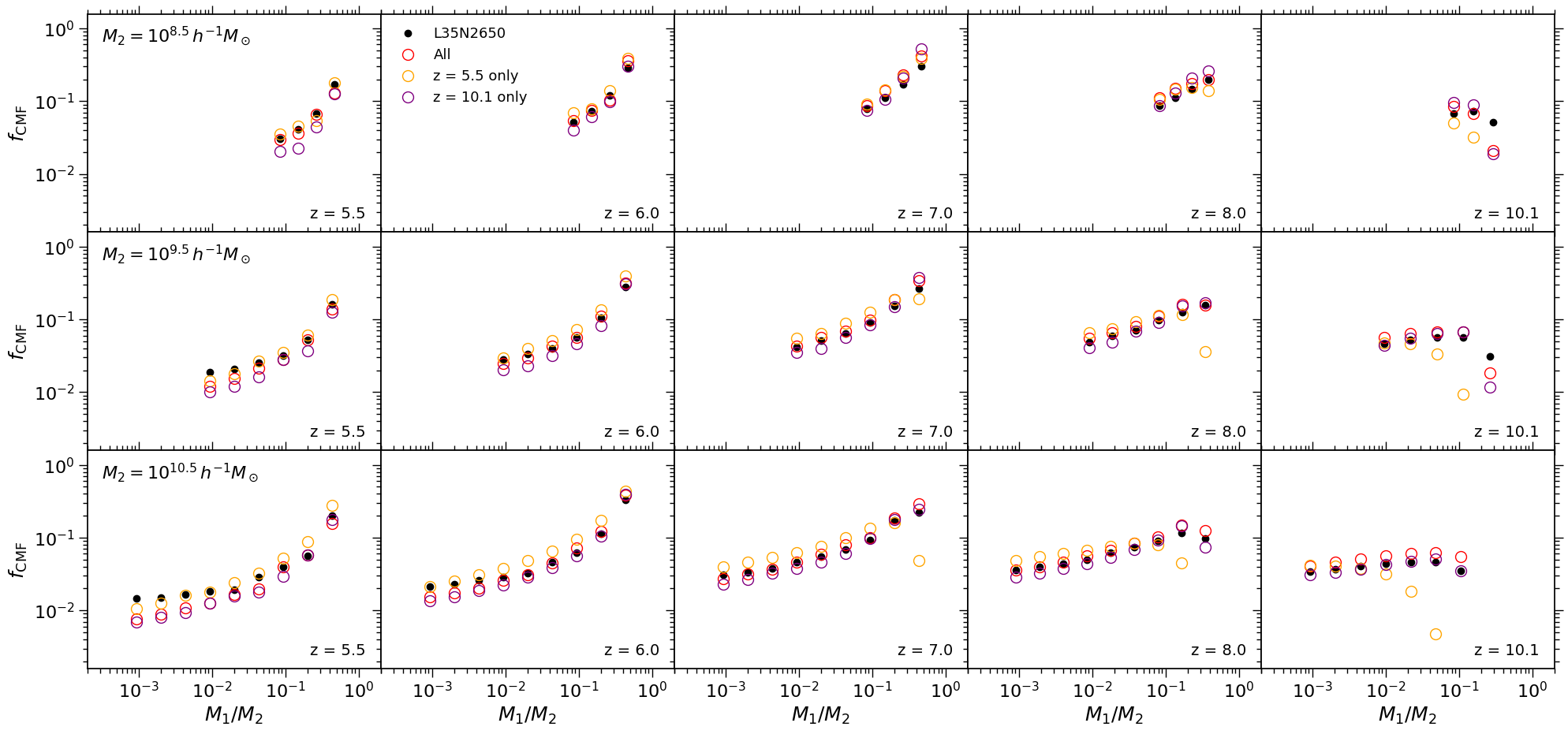}
    \caption{Fitting results of two additional calibrations for the \citet{2008MNRAS.383..557P} algorithm. The conditional mass functions are defined by $d f_\text{CMF} / d \ln M_1$. Black dots are the fitting data, which are estimated using L35N2650. Red empty circles are the same as those in Figure \ref{fig:fit}. Yellow and purple empty circles are the results that use $z = 5.5$ data only and $z = 10.1$ data only respectively. Their corresponding parameters are listed in Table \ref{table:params_mc_sp}.}
    \label{fig:fit_sp}
\end{figure*}

In Section \ref{sec:mc_tree}, we do not employ any weights for different mass and redshift ranges in the cost function for calibrating the \cite{2008MNRAS.383..557P} algorithm. In this appendix, we present two additional calibrations of the algorithm to show the potential bias of this treatment. In Figure \ref{fig:fit_sp}, the calibration results that use $z = 5.5$ data only and $z = 10.1$ data only are shown as yellow and purple empty circles respectively. The corresponding parameters are listed in Table \ref{table:params_mc_sp}. The result that uses $z = 5.5$ data only is improved at $z = 5.5$ but becomes significantly poorer at higher redshifts. In terms of the purple empty circles, the fitting is improved at $M_2 = 10^{10.5} \hunit M_\odot$, $z = 10.1$ and is similar or slightly poorer at other mass and redshift ranges. These results suggest that the employment of weighting may only provide moderate improvements on the calibration of the \cite{2008MNRAS.383..557P} algorithm, which, however, is purely artificial. Therefore, we do not employ any weights on the calibration and adopt the parameters obtained in Section \ref{sec:mc_tree} as the fiducial model in this work.

\begin{table} 
    \centering
	\caption{Results of two additional calibrations for the \citet{2008MNRAS.383..557P} algorithm.}
    \label{table:params_mc_sp}
    \begin{tabular}{c|c|c|c}
	\hline \hline
    Symbol & All & $z = 5.5$ only  & $z = 10.1$ only \\
	\hline
    $G_0$      & 1.0 & 0.7 & 0.6 \\
    $\gamma_1$ & 0.2 & 0.2 & 0.5 \\
    $\gamma_2$ & -0.4 & 0.4 & -0.1 \\
    \hline
    \end{tabular}
\end{table}

\bsp
\label{lastpage}
\end{document}